\documentclass[]{spie}  

\usepackage{amsmath,amsfonts,amssymb}
\usepackage{graphicx}
\usepackage[colorlinks=true, allcolors=blue]{hyperref}
\usepackage{wrapfig}

\usepackage{placeins}
\usepackage{caption}
\usepackage{subcaption}
\usepackage{graphbox}
\usepackage[export]{adjustbox}
\usepackage{gensymb}
\usepackage{amsmath}
\usepackage{wrapfig}

\title{ Manifold Reconstruction of Differences: A Model-Based Iterative Statistical Estimation Algorithm With a Data-Driven Prior \vspace{-1mm}}
\author[]{\normalsize\vspace{-3mm} Matthew Tivnan, J. Webster Stayman, \em Biomedical Engineering, Johns Hopkins University, Baltimore, MD\vspace{-3mm}}

\begin{document} 
\maketitle
\vspace{-6mm}
\begin{abstract}
\vspace{-2mm}
Manifold learning using deep neural networks been shown to be an effective tool for building sophisticated prior image models that can be applied to noise reduction in low-dose CT. 
We propose a new iterative CT reconstruction algorithm, called Manifold Reconstruction of Differences (MRoD), which combines physical and statistical models with a data-driven prior based on manifold learning. The MRoD algorithm involves estimating a \textit{manifold component}, approximating common features among all patients, and the \textit{difference component} which has the freedom to fit the measured data. By applying a sparsity-promoting penalty to the difference image rather than a hard constraint to the manifold, the MRoD algorithm is able to reconstruct features which are not present in the training data. The difference component itself may be independently useful. While the manifold captures typical patient features (e.g. healthy anatomy), the difference image highlights patient-specific elements (e.g. pathology). In this work, we present a description of an optimization framework which combines trained manifold-based modules with physical modules.
We present a simulation study using anthropomorphic lung data showing that the MRoD algorithm can both isolate differences between a particular patient and the typical distribution, but also provide significant noise reduction with less bias than a typical penalized likelihood estimator in composite manifold plus difference reconstructions. 
\end{abstract}


\vspace{-5mm}
\section{INTRODUCTION}

\vspace{-3mm}

Manifold learning is a form of data-driven nonlinear dimensionality reduction which has shown great promise in the task of low-dose CT reconstruction. Such learning involves finding a latent representation of the image data in a lower-dimensional space than the original voxelized basis. This latent space, or manifold, can encode for a large portion of the variations within a certain class of images, and is coupled with a nonlinear mapping from the manifold back to the original image space. The transformation between the image space and the latent space is trained using, e.g., deep neural networks (DNN) \cite{bengio2006nonlocal}. A convolutional architecture has found use in image denoising\cite{hein2007manifold}, and has also been incorporated into full deep learning CT reconstruction models \cite{zhu2018image}. 
One of the major concerns regarding deep learning CT reconstruction is that the processing may be highly object-dependent and can have a highly nonlinear response to specific stimuli\cite{gang2019generalized}. Moreover, there is also a risk of over-fitting to the training data. For example, if a new patient has a lesion that was not represented in the training data, it may be difficult or impossible to accurately represent that feature if the image result is constrained to lie on the manifold. The net effect is that certain features that are present may not be visualized in the data, or vice-versa, false features may be injected into the image, but the overall image has the appearance of ``good'' image quality. Another critique of this type of hard constraint to the manifold is that there is no explicit check to ensure goodness of fit between image estimates and measured data. However, some efforts have been made to combine manifold-based noise reduction with a physics-based statistical estimators by alternating between a maximum likelihood reconstruction and a DNN-autoencoder-based denoising module \cite{ma2018low}.

We propose a new iterative CT reconstruction algorithm, called Manifold Reconstruction of Differences (MRoD), which estimates images using a physics-based statistical data-fidelity term, and a manifold-based penalty term. These terms are combined into one penalized-likelihood estimator which is optimized directly. The MRoD algorithm involves jointly estimating two images: the \textit{manifold component}, which contains features which are typical among all patients, and the \textit{difference component} which contains the changes needed to fit the measured data. A final composite image estimate may be formed via sum of these two components. A penalty is applied to minimize the difference component and promote sparsity. This effectively encourages image estimates which very close to manifold-constrained case without invoking a hard constraint. These small changes in the difference image are guided by a data-fidelity term which uses physics-based signal and noise models to evaluate the goodness-of-fit between the final images (a summation the two components) and the measured data. Therefore, the MRoD algorithm has the ability to reconstruct image features which may be rare or irregular and did not appear in the manifold training cases. 
Difference images themselves may independent, clinically useful outputs. For example, while the manifold component captures features which are typical across all patients (e.g. healthy anatomy), the difference highlights patient-specific atypical elements (e.g. pathology or rare variations within the patient population). It is often these differences that are of greatest interest to radiologists.

In this work, we present a mathematical description of the algorithm as well a description of the software implementation. This includes the combination of trained deep DNN modules, forward/back-projectors, and other physics-based modules in a unified iterative optimization framework.
In a simulated lung imaging study, we characterize the effect of the penalty weight on noise and bias. Comparisons to maximum-likelihood and roughness-penalized likelihood estimators are also provided. Finally, we present a numerical experiment which involves reconstructing pathological cases. This demonstrates that the MRoD algorithm is capable of imaging clinically meaningful irregularities which were not present in the training data.

\vspace{-4mm}
\section{METHODS}
\vspace{-1mm}
\subsection{Manifold Learning with a DNN Encoder-Decoder}

\vspace{-2mm}

\begin{figure}
    \centering
    \includegraphics[width=0.95\textwidth]{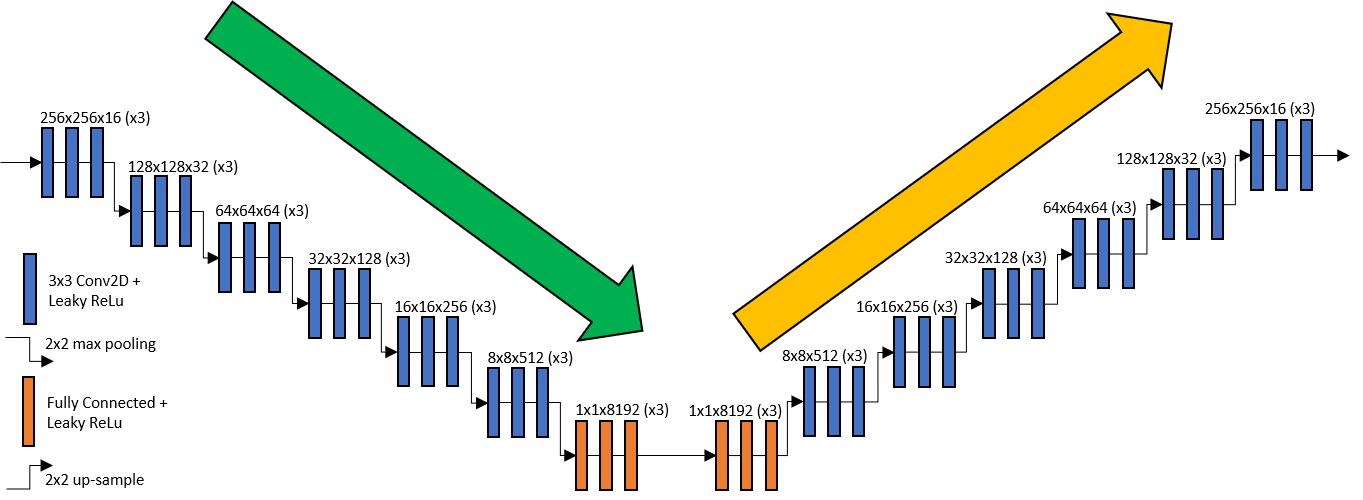}
    \vspace{5mm}
    \caption{The DNN encoder-decoder model which is used for manifold learning. The encoder portion (green arrow) transforms a 256x256x1 image of to a 1x1x8192 manifold representation and the decoder portion (yellow arrow) transforms the manifold back to the image space. For each training sample, the same image is provided as both input and output to the model so this lossy nonlinear compressor can be trained to extract the most important image features. \vspace{-5mm}}
    \label{fig:EncoderDecoder}
\end{figure}

We consider a DNN encoder-decoder model consisting of an encoder which compresses the image data to the manifold representation and a decoder which maps that manifold back to the original image space. If the $N$ voxels comprising the image data are represented as a single vector $\mathbf{x} \in \Re^N$, then the encoder is a nonlinear operation $\mathbf{m} = \mathcal{E}(\mathbf{x}) \in \Re^M$ which maps the image data onto an $M$-dimensional manifold where $M < N$. The decoder is another nonlinear operation $\Tilde{\mathbf{x}} = \mathcal{D}(\mathbf{m}) \in \Re^N$ which maps the manifold back to the image space. Trained weights for the DNN are optimized to minimize the L1-norm of the residual, $| \mathbf{x} - \mathcal{D}(\mathcal{E}(\mathbf{x}))|$.
Fig. \ref{fig:EncoderDecoder} shows the DNN encoder-decoder consisting of convolutional and fully connected layers with leaky ReLu activation functions. The input and output images have 256x256x1 voxels and the intermediate manifold representation has 1x1x8192 features. The encoder has 2x2 max pooling layers and the decoder has 2x2 up-sampling layers. In the following sections, we will incorporate this data-driven manifold as a penalty in an iterative CT reconstruction algorithm to encourage image features which were common in the training data.


\vspace{-3mm}

\subsection{Penalized-Likelihood Estimation Framework}
\vspace{-2mm}

The MRoD algorithm can be described mathematically as a penalized-likelihood estimator which includes a physics-based data-fidelity term and a manifold-based penalty. A general physical model for x-ray CT measurements is given by  $\mathbf{\bar{y}}(\mathbf{x})=\mathbf{B}\exp(-\mathbf{A}\mathbf{x})$, where $\mathbf{A}$ is a forward projector used to model the system geometry and $\mathbf{B}$ is a diagonal scaling operator used to model the gain. A penalized weighted least-squares objective function may be written as
\vspace{-3mm}
\begin{gather}
    \Phi(\mathbf{x}) = L(\mathbf{x} | \mathbf{y}) + R(\mathbf{x}),  \quad  \quad \quad L(\mathbf{x} | \mathbf{y}) = (\mathbf{y} - \mathbf{\bar{y}}(\mathbf{x}))^T \mathbf{\Sigma_y^{-1}}(\mathbf{y} - \mathbf{\bar{y}}(\mathbf{x}))\vspace{-9mm}
\end{gather}
\vspace{-8mm}

\noindent
The objective function $\Phi(\mathbf{x})$ has a weighted least-squares term, $L(\mathbf{x}|\mathbf{y})$, which quantifies the data fidelity with respect to the measured CT data, $\mathbf{y}$, and a penalty term $R(\mathbf{x})$ which serves as a regularizer. In this work, we will compare the new proposed manifold-regularized case with a standard quadratic roughness penalty given by $R(\mathbf{x}) = \beta \mathbf{x}^T \mathbf{R} \mathbf{x}$ where $\mathbf{R}$ is a discrete Laplace operator and $\beta$ is the penalty strength. 


\vspace{-3mm}
\subsection{Manifold Reconstruction of Differences}
\vspace{-2mm}

The MRoD algorithm is inspired by the Reconstruction of Difference (RoD) algorithm \cite{pourmorteza2016reconstruction}
which involves estimating the difference with respect to some reference image (e.g. from a previous scan). However, the MRoD algorithm does not require an \textit{a priori} reference image. Instead, the reference image is replaced with a manifold-based approximation image which is jointly estimated together with the difference component. MRoD is a penalized-likelihood estimator with an L1-norm penalty to promote small sparse difference estimates. The proposed estimator can be formulated as follows.

\vspace{-8mm}

\begin{gather}
    \Phi(\mathbf{m}, \mathbf{x}_D) = L(\mathcal{D}(\mathbf{m}) + \mathbf{x}_D | \mathbf{y}) + \gamma |\mathbf{x}_D|, \quad   \quad \quad \hat{\mathbf{m}}, \hat{\mathbf{x}}_D = \underset{\mathbf{m}, \mathbf{x}_D}{\text{argmin}} \enspace \Phi(\mathbf{m}, \mathbf{x}_D), \quad   \quad \quad \hat{\mathbf{x}} = \mathcal{D}(\hat{\mathbf{m}}) + \hat{\mathbf{x}}_D
\end{gather}
\vspace{-7mm}

\noindent The nonlinear operator $\mathcal{D}$ is the decoder portion of the trained encoder-decoder model. Therefore, the manifold component, $\mathcal{D}(\hat{\mathbf{m}})$, is an approximate image which lies on the manifold. The likelihood term is evaluated using the final composite image estimate,  $\mathbf{\hat{x}} =  \mathcal{D}(\mathbf{\hat{m}}) + \mathbf{\hat{x}}_D$, where $\hat{\mathbf{x}}_D$ is the difference component. The regularization strength is controlled by the parameter $\gamma$. As $\gamma$ approaches zero, the difference image is unconstrained so the estimator approaches a maximum-likelihood estimator. As $\gamma$ approaches infinity, the the solution will be constrained to lie on the manifold. 

\vspace{-3mm}
\subsection{Lung Imaging Simulation Study}
\vspace{-2mm}

To demonstrate the MRoD algorithm, we conducted a simulated lung imaging study. Data were generated using the XCAT 2.0 anthropomorphic digital phantom. The phantom was parameterized by 97 values which represent volumes and sizes of various organs and other anatomical structures as well as stretches and rotations of the entire body. For each realization, the patient had a 50\% chance to be male or female, and the remaining parameters were randomly generated according to a multivariate Gaussian distribution with a standard deviation of 3\% for size/volume parameters, 2 degrees for rotation parameters, and a correlation coefficient of 0.75 between size/volume parameters and 0.00 for all other combinations. These parameters were initialized for 100 patients in the training data. Digital phantoms were $256\times256\times100$ voxels with voxel size $2\times2\times2$~mm. Voxel values represented attenuation coefficients at 100 keV. Image volumes were cropped along the inferior-superior axis to include slices between the diaphragm and the superior extent of the lungs. Slices were randomly shuffled across all data since training and data processing was conducted on a 2D slice-by-slice basis. The DNN encoder-decoder shown in Figure \ref{fig:EncoderDecoder} was trained by using each training sample as a $256\times256\times1$ input and output with a mean absolute error loss function. The trained decoder was then used as the $\mathcal{D}$ operator as described in a previous section. Synthetic data were generated based on circular fan-beam CT system with 1100~mm source-to-detector distance, 830~mm source-to-axis distance, and 1~mm detector pixels. Uncorrelated Poisson noise was added to the synthetic measured data according to a dose of $10^5$ photons per pixel per view.

Both MRoD and QPL estimators were implemented using the Keras deep learning framework \cite{chollet2015keras}. Custom static layers were designed for forward- and back-projectors, and other operations needed for the ``physics-based'' modules. Thus, iterative reconstruction uses the same optimization framework and software that is typically used to train a DNN. The only variable quantities during optimization were the image and manifold estimates. For the MRoD case, the pre-trained decoder was included in the estimation model, but the weights are treated as fixed during the iterative CT reconstruction. The estimation was conducted using 1,000 iterations of the Adam optimizer with a learning rate of 0.001. 

 Three additional simulated patients (distinct from training data) were generated using the above process and a slice at the center of the lungs was extracted for evaluation. The first image was used to compare the noise-bias relationship for the MRoD and QPL algorithms by conducting a sweep of the penalty strength parameters $\beta$ and $\gamma$. A noiseless version of the data was also generated for this case. Bias is defined as the Root-Mean-Squared Error (RMSE) between the image estimates for the noiseless case and the ground truth, and noise is defined as the RMSE between estimates for the noisy and noiseless cases. The second and third images are used to simulate pathological cases. For the second image, we manually added a spiculated lung nodule, and for the third image we manually added a 4mm crack in a rib. These pathological features were not present in the training data.

\vspace{-4mm}
\section{RESULTS}
\vspace{-2mm}

\begin{figure}[h!]
    \centering
    \includegraphics[width=0.99\textwidth]{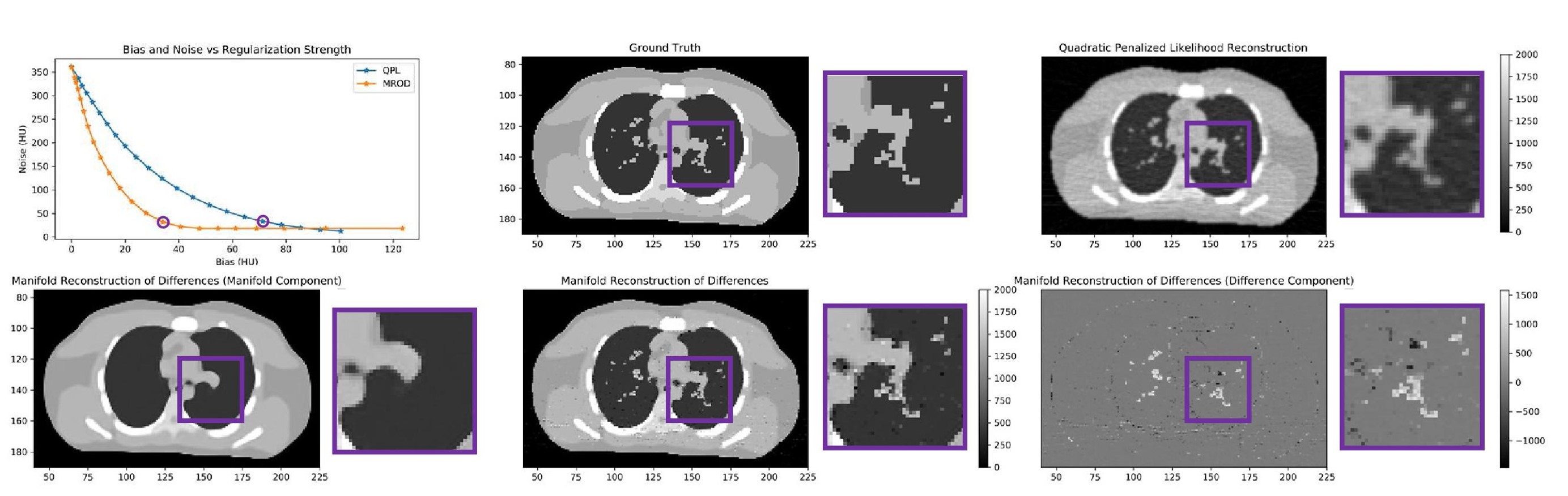}
    \caption{ Noise vs bias and example reconstructions for QPL and MRoD. Violet circles indicates image examples shown.  }
    \label{fig:NoiseVsBias}
\end{figure}

The results from the first study are shown in Figure \ref{fig:NoiseVsBias}. Each point on the noise vs bias plot represents a different reconstruction for which the level of noise and bias were evaluated. For low penalty strengths, the two algorithms have very similar behavior, since the limiting behavior for both algorithms is a maximum-likelihood estimator. However, for the range of practically useful penalty strengths, the MRoD algorithm has significantly less bias for a matched noise level. We select the case for each method with a noise level closest to 30~HU, which corresponds to relatively heavy regularization, and show the image estimates alongside the ground truth. As expected, QPL bias appears as spatial blur. For the MRoD case, we show the manifold component, the difference component, and the final image estimate. The manifold component captures the general shape of the patient and consists of extremely smooth regions since it was trained with noiseless ground-truth phantoms. Much soft tissue variation in the lungs is absent from the manifold component. 
However, such variations are represented in the difference image where the data-fidelity term of the objective function may compensate for this residual density in the lungs. The final MRoD image appears to be a close match to the ground truth with much less bias due to spatial blur than the QPL image.

\begin{figure}[h!]
    \centering
    \includegraphics[width=0.99\textwidth]{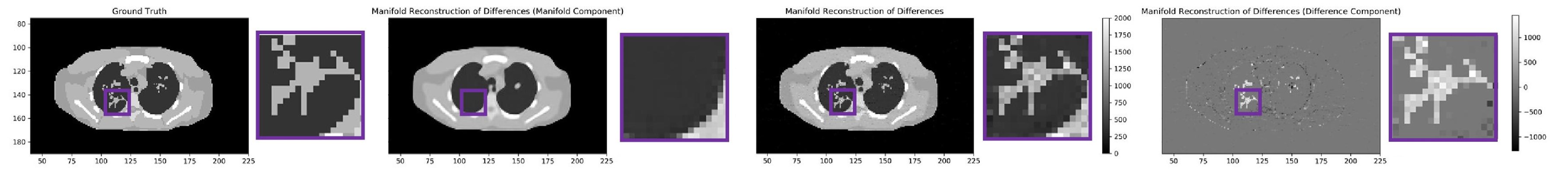}
    \caption{ Spiculated lung nodule example. Ground truth, manifold, difference, and final images are shown. }
    \label{fig:nodule}
\end{figure}

\begin{figure}[h!]
    \centering
    \includegraphics[width=0.99\textwidth]{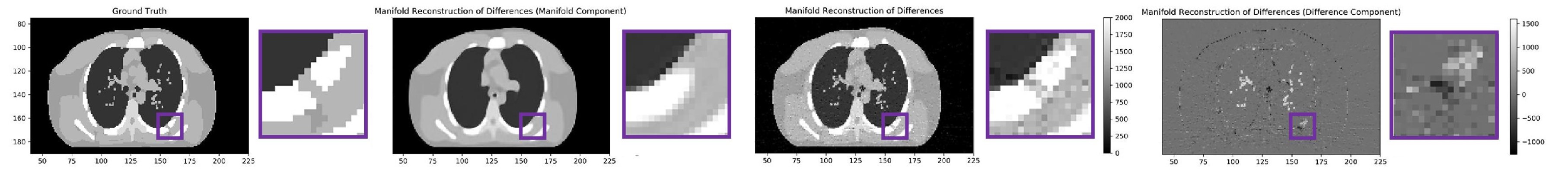}
    \caption{ Cracked rib example. Ground truth, manifold, difference, and final images are shown. }
    \label{fig:rib}
\end{figure}

Figures \ref{fig:nodule} and \ref{fig:rib} show the ground truth and MRoD reconstructions of the pathological cases for the lung nodule and the cracked rib. In both cases, the manifold component shows an approximation which does not capture these features entirely since they did not appear in the training data. The manifold component for the nodule is empty space. The manifold component for the cracked rib looks like an intact rib that extends through the cracked region. This is the danger of over-fitting to the training data faced by many deep learning CT reconstruction algorithms. However, the MRoD algorithm is guided by physics-based models to compensate for these errors in the difference image. For the nodule case, the spiculated lesion is clearly shown in the difference image. For the cracked rib case, the difference component contains both the negative signal needed to represent the crack, as well as the positive signal needed to represent the full extent of the rib. In both cases, the final MRoD image estimate appears to be a close match to the ground truth in a diagnostically useful way. This is promising because it indicates that the MRoD algorithm is capable of both representing features that were not present in the training data, but also highlighting the presence of these differences from the manifold of typical patients.

\FloatBarrier

\vspace{-3mm}
\section{CONCLUSION}
\vspace{-2mm}

The MRoD algorithm combines advantages of physics-based statistical estimation and deep-learning CT. It uses sophisticated physical models, statistical weights, and prior knowledge about the estimates via the manifold-based penalty. Unlike many other deep learning CT methods, the final image estimates are not constrained lie on the manifold. The estimates are guided by physics-based models which allow for deviation from prior expectations when needed. 
The experimental results show that the MRoD algorithm is highly effective at reducing noise with small levels of bias. The results from from the pathological examples show that this method is capable of reconstructing image features that never appeared in the training data. Finally, both the composite manifold plus different images and difference component alone appear to be clinically useful outputs, where the latter element can highlight differences from typical anatomy showing what is unique about this patient. 

\acknowledgments 
The authors thank Dr. Paul Segar who provided access to the XCAT 2.0 anthropomorphic digital phantom software.
This work is supported, in part, by NIH grant R21CA219608.
\bibliography{report}{}
\bibliographystyle{spiebib-abbr} 

\end{document}